\documentclass[12pt,preprint]{aastex}
\def\msol{{\rm M}_\odot}
\def\msolyr{{\rm M}_\odot {\rm yr^{-1}}}
\def\simle{\,\hbox{\hbox{$ < $}\kern -0.8em \lower 1.0ex\hbox{$\sim$}}\,}
\def\te{T_{\rm eff}}
\shortauthors{THORSTENSEN ET AL.}
\shorttitle{64-minute Dwarf Nova}
\begin{document}
\title{1RXS J232953.9+062814: A Dwarf Nova with a 64-minute Orbital 
Period and a Conspicuous Secondary Star
\footnote{Based in part on
observations obtained at the MDM Observatory, operated by
Dartmouth College, Columbia University, Ohio State University, and
the University of Michigan.}
}

\author{John R. Thorstensen, William H. Fenton}
\affil{Department of Physics and Astronomy\\
6127 Wilder Laboratory, Dartmouth College\\
Hanover, NH 03755-3528;\\
john.thorstensen@dartmouth.edu, w.h.fenton@dartmouth.edu}
\author{Joseph O. Patterson, Jonathan Kemp\altaffilmark{2}}
\affil{Department of Astronomy, Columbia University\\
538 West 120th Street, New York, NY 10027;\\
jop@astro.columbia.edu, j.kemp@jach.hawaii.edu}
\author{Thomas Krajci}
\affil{Center for Backyard Astrophysics (New Mexico), 
1688 Cross Bow Circle, Clovis, NM 88101;\\
krajcit@3lefties.com}
\author{Isabelle Baraffe}
\affil{\'Ecole Normale Sup\'erieure, 69364 Lyon Cedex 07, France;\\
ibaraffe@ens-lyon.fr}
\altaffiltext{2}{Also at Joint Astronomy Center, Hilo, Hawaii.}

\begin{abstract}
We present spectroscopy and time-series photometry of 
the newly discovered dwarf nova 1RXS J232953.9+062814.  Photometry
in superoutburst reveals a superhump with a period of 
66.06(6) minutes.  The low state spectrum shows Balmer
and HeI emission on a blue continuum, and in addition shows
a rich absorption spectrum of type K4 $\pm$ 2.
The absorption velocity is modulated sinusoidally
at $P_{\rm orb} = 64.176(5)$ min, with semi-amplitude
$K = 348(4)$ km s$^{-1}$.  The low-state light curve is double-humped
at this period, and phased as expected for ellipsoidal
variations.  The absorption strength
does not vary appreciably around the orbit. 
The orbital period is shorter than any other
cataclysmic variable save for a handful of helium-star
systems and V485 Centauri (59 minutes).  
The secondary is much hotter than main sequence stars of 
similar mass, but is well-matched by helium-enriched models,
indicating that the secondary evolved from a more massive
progenitor.  A preliminary calculation in which a 1.2 M$_{\odot}$ 
star begins mass transfer near the end of H burning
matches this system's characteristics remarkably well.

\end{abstract}

\section{Introduction}

Cataclysmic variable stars (CVs) are close binary systems
in which a low-mass secondary transfers mass onto a white dwarf;
\citet{warn} wrote an excellent monograph on CVs.
Orbital angular momentum losses $\dot J$ evidently drive
CV evolution.  As the orbit shrinks, the secondary star's
Roche critical lobe contracts, causing mass transfer.  
The Roche geometry tightly constrains the secondary star's mass
at a given orbital period $P_{\rm orb}$.  Short-period systems
have low-mass secondaries, so if the chemical composition  
is normal ($X \sim 0.7$), 
the secondary is faint and contributes negligibly
to the visible-light spectrum (Fig.~4 of \citealt{patprecess01}).
For normal compositions the radius is expected
to reach a minimum around 0.05 $M_{\odot}$, leading to a predicted
period minimum around 70 -- 75 minutes (the exact value being dependent
on $\dot J$), with subsequent evolution
driving the system to greater separations
\citep{kb99,patlate98,ps81}.  

During outburst, 
short-period dwarf nova systems often show photometric oscillations
(superhumps) at periods a few percent longer than $P_{\rm orb}$.  
The superhump frequency is thought to be the beat between 
$P_{\rm orb}$ and a tidally driven precession
of an eccentric disk.  The fractional period excess 
of the superhump appears to be a measure of the mass ratio 
\citep{patprecess01}. 

We recently observed 1RXS J232953.9+062814 (hereafter
RX2329+06), a newly-recognized dwarf 
nova system with $P_{\rm orb}$ below the
canonical minimum period for hydrogen-rich secondary stars.
Our observations reveal an unexpectedly hot secondary star.
We suggest that the secondary has undergone substantial nuclear
evolution leading to an enhanced helium abundance.  

\section{Observations and Analysis}

Our optical photometry is from differential CCD 
measurements, mostly from the Center for Backyard
Astrophysics (CBA) worldwide network of small telescopes \citep{cba93}.  
Our spectroscopy is from the 2.4 m 
Hiltner telescope at MDM Observatory on Kitt Peak.  
Details of the instrumental
setup and reduction were as described in \citet{t98}.
We obtained a total of 77 exposures of 360 s each, on 
2001 Nov 20 - 22 and 24 UT, spanning 6.0 h of hour angle.

We defer detailed discussion of the outburst photometry
for a longer paper.  Briefly, the superhump period during
outburst was 0.04637(4) d, decreasing gradually over the
next 10 days to around 0.0458 d.  

For astrometry, we obtained several $I$-band images with the 
1.3m McGraw-Hill telescope at MDM.  We compared 
these with archival sky-survey data from the USNO Flagstaff
Station Image and Catalogue Archive, and found 
$\mu = 61 \pm 11$ mas yr$^{-1}$ in
P.A. $331 \pm 10$ degrees.  The ICRS position at
J2000, referred to USNO A2.0 \citep{mon96} and five Tycho-2
stars \citep{Hog2000}, was $\alpha = 23^{\rm h} 29^{\rm m} 54^{\rm s}.22$, 
$\delta = +6^{\circ} 28' 11''.8$, $\pm 0''.3$

Fig.~1 shows the mean low-state spectrum.  Broad
Balmer emission is prominent, along with weaker emission
from He I (Table 1).  The Balmer lines show double peaks most of 
the time, separated by $1000 \pm 100$ km s$^{-1}$ when the
peaks are most distinct.  The continuum has $F_{\lambda} \propto 
\lambda^{-1.6}$, and the flux level indicates $V = 16.2 
\pm 0.3$ (estimated uncertainty).  
The spectrum also shows conspicuous absorption features
which appear consistent with a late-type star.

We measured absorption velocities by cross-correlating
our spectra against a rest-frame sum of several IAU velocity
standards, using {\it xcsao} \citep{kurtzmink98}.
Although the absorption was nearly invisible in the
individual spectra, the cross-correlation technique succeeded
in all cases, giving typical formal errors $\sim
15$ km s$^{-1}$.  A least-squares period search of the absorption 
velocities \citep{tpst} showed a strong, unambiguous period near
64 minutes, and a sine fit (Fig.~2) gave the parameters in
Table 2.  H$\alpha$ emission-line velocities measured
using a convolution method sensitive to the line wings
\citep{sy} gave noisier velocities (also in Fig.~2), 
but yielded the same period.
Fig.~3 shows a greyscale representation of the spectra 
with phase \citep{tay98}.  The absorption spectrum and 
its modulation are obvious, and the H$\alpha$ emission line 
also shows an apparent $S$-wave.

Using library spectra from \citet{pickles98}, we estimate
the secondary's spectral type to be K4 $\pm 2$ subclasses.
Earlier spectral types do not show CaI $\lambda$6162 as strongly
as RX2329+06, and later types begin to show
molecular features which are not present here. 
Assuming the secondary's absorption line strengths are 
similar to the library spectra,
we estimate that the secondary contributes 
$50 \pm 20$ per cent of the light near 5500 \AA . 

The bottom panel of Fig.~2 shows phase-averaged 
white-light differential photometry from the CBA network.
The double-peaked modulation is consistent with ellipsoidal
variations, which are not unexpected since the secondary
contributes so strongly to the total light.

\section{Inferences}

{\it Distance.} The secondary spectrum and period
constrain the distance.  
From Roche lobe arguments we infer $R_2 = 0.12 \pm 0.03$ R$_{\odot}$ 
(see below), where the uncertainty is conservative since 
$R_2$ scales as $M_2^{1/3}$.
K-star surface brightnesses inferred from Table
3 of \citet{beuermann99} then imply that the secondary
has $M_V = 10.4 \pm 0.7$. Our data give $V = 17.0 \pm 0.5$ for
the secondary alone, where the uncertainty includes 
the flux calibration and the secondary's relative
contribution.  The distance is then
210 (+110, $-70$) pc.  At the nominal distance the 
transverse velocity is 61 km s$^{-1}$.
The quiescent disk at 210 pc has 
$M_V \sim 10$, rising to $M_V \sim 5$ in outburst,
which agrees well with 
the $M_V$-$P_{\rm orb}$ relation of \citet{warn87}.

{\it Masses.} There are no eclipses, so the inclination $i$
is uncertain.  The secondary's
velocity amplitude gives a mass function of 0.194(6) M$_{\odot}$.  
This may be distorted by illumination effects, but we see 
no evidence of variation of the line features around the spectrum.
The secondary is 
presumably at least $0.05$ M$_{\odot}$, which implies a white 
dwarf mass $M_1 > 0.28$ M$_{\odot}$,
not a particularly stringent constraint.
In principle, the ellipsoidal variations could
constrain $i$, but this is 
sensitive to the assumed limb-darkening parameters
\citep{boch79} and to the fractional contribution from the 
secondary.  We tried measuring the emission lines with 
a range of convolution parameters \citep{sha83}, but found
a consistent 0.05-cycle phase offset between the emission
velocities and the expected phase of the white dwarf motion,
so we do not use the emission lines to estimate the mass ratio. 

The mass ratio can be estimated from the 
superhump period excess $\epsilon = 
(P_{\rm sh} - P_{\rm orb}) / P_{\rm orb}$,
using the $\epsilon$-$q$ relationship calibrated by 
\citet{patprecess01}.  At maximum light, 
$\epsilon = 0.040$, implying $q = M_2/M_1 = 0.185$,
or $M_2 = 0.13$ M$_{\odot}$ for
a broadly typical $M_1 = 0.7$ M$_{\odot}$.
The Roche constraint implies $R_2/R_{\odot} = 
0.245 (M_2 / M_{\odot})^{1/3}$ at this period \citep{beuermann98};
combining this with $q$ yields $R_2 = 0.124 (M_1/0.7 M_{\odot})^{1/3}$.
The illustrative system with $(M1,M2) = (0.7,0.13)$ M$_{\odot}$ would
reproduce the observed mass function at 
$i = 47$ degrees.

\section{Discussion}

It is surprising to find such a hot secondary at this short period,
where the secondary mass must be $\sim 0.1$ M$_{\odot}$.
No calculation beginning with a solar-abundance 
secondary has produced periods this short \citep{kb99}, and 
hydrogen-rich stars of this small a mass would be much
cooler than observed.

We suggest that the secondary is substantially enriched in helium, as
a result of nuclear evolution prior to mass transfer. 
As shown by \citet{bk00}, secondary donors 
which have evolved off the Zero Age Main Sequence (ZAMS) 
at the onset of mass transfer
can explain a substantial fraction of the observed CVs with
late spectral types and $P_{\rm orb} > 6$ h.
Test calculations with constant mass transfer rates
show that significantly evolved donors (e.g. with mass transfer starting near
the end of central H burning) can be much hotter at a given $P_{\rm orb} 
\simle 5$ h than ZAMS donors. This is illustrated in
Fig. \ref{fig4}, which displays sequences with initial donor mass
$M_2$ = 1.2 $\msol$ and starting mass transfer on the ZAMS (solid line)
or near the end of H burning (dashed line).
As already noted by \citet{bk00}, such extreme sequences never
become fully convective, because of the lower central H abundance, 
and may continue to transfer mass in the 2-3 h period gap. 
If they sequences do so, they can reach very low orbital 
periods with unusually high $\te$. 

For the test case shown in Fig.~\ref{fig4}, the evolved sequence 
(dashed line) reaches $P_{\rm orb}$ = 64 min with a mass 
$M_2 \sim$ 0.11 $\msol$ and
a radius $R_2$ $\sim$ 0.13 $R_\odot$, in close agreement with the estimates
in \S 3. We derive a spectral type K5, based on the empirical SpT - $(I-K)$ 
relation of Beuermann et al (1998), again matching observation.
We stress that our spectral type estimate is to be treated with caution,
since the surface chemical composition is expected to be non-standard,
with a mass fraction of H  $\sim$ 30-40\%. Because  
CNO-processed material should be visible on the surface,
the model has C significantly depleted and N enriched by a factor 
$\sim$ 5, whereas O is hardly affected.

We stress that this evolutionary sequence is preliminary,
but it provides a surprisingly good match to the observed 
properties of RX2329.

The emission line fluxes (Table 1) suggest that helium is enhanced.
The ratio of H$\alpha$ to HeI $\lambda6678$ is
about 3.6.  We measured the H$\alpha$/$\lambda$6678 ratio in 
archival spectra of SU UMa stars \citep{tpst,tt97,thor97} 
and found typical values of 8, with none below 6.  This could be an
excitation effect, but even high-excitation novalikes such
as V603 Aql \citep{patt97} have larger ratios.  The 59-minute
binary V485 Cen also appears to have an unusually low
H$\alpha$/$\lambda$6678 in the spectrum published by
\citet{aug96}, but absorption features are not evident there.
As noted earlier, the relative CNO abundances
should be affected, but we cannot comment on this since CNO elements
lack strong lines at this $\te$.

In sum, RX2329 is a CV in which the secondary evidently has 
undergone significant nuclear burning and then had much of
its mass stripped away, since at its present mass, its nuclear 
evolution timescale greatly exceeds the Hubble time.
If the white dwarf mass is similar to most CVs ($\sim 0.7$ M$_{\odot}$),
much of the mass lost by the secondary appears to have been  
lost from the system.  Evidently
the system formed with a secondary considerably
more massive than the white dwarf.  In principle this leads
to mass transfer on a thermal timescale, until $M_2$ 
becomes small enough for the system to reappear as a standard
CV (see \citealt{bk00} for a discussion).

{\it Acknowledgments.} We gratefully acknowledge funding by the NSF 
(AST 9987334), and we thank the  
MDM staff for their excellent support.  Special thanks go to the 
CBA observers who contributed to the photometry, including
Dave Skillman, Arto Oksanen, Ed Beshore, and Tonny Vanmunster; 
a fuller report on their work will be forthcoming.  We made 
use of the USNOFS Image and Catalogue Archive
operated by the United States Naval Observatory, Flagstaff Station
(http://www.nofs.navy.mil/data/fchpix/).

{\it Note added 2002 January 29:  We obtained seven more spectra with the 2.4m on
2002 Jan. 21 UT.  Combining absorption velocities from these with the
2001 November data yields a refined $P_{\rm orb}$ = 0.0445671(2) d.}

\clearpage

\begin{deluxetable}{lrcc}
\tablewidth{0pt}
\tablecolumns{6}
\tablecaption{Emission Features}
\tablehead{
\colhead{Feature} & 
\colhead{E.W.\tablenotemark{a}} & 
\colhead{Flux\tablenotemark{b}}  & 
\colhead{FWHM \tablenotemark{c}} \\
 & 
\colhead{(\AA )} & 
\colhead{(10$^{-15}$ erg cm$^{-2}$ s$^{1}$)} &
\colhead{(\AA)} \\
}
\startdata
H$\gamma$  & 12  & 30 & 32  \\
HeI$\lambda$4471  & 4:  & 9: & 24:  \\
HeII$\lambda$4686 & 3: & 5: & 25: \\
H$\beta$ & 18 & 29 & 32 \\
H$\alpha$ & 25 & 26 & 41 \\
HeI $\lambda$6678  & 7  & 7 & 49 \\
HeI $\lambda$7067  & 7  & 6 & 54 \\
\enddata
\tablenotetext{a}{Emission equivalent widths are counted as positive.}
\tablenotetext{b}{Absolute line fluxes are uncertain by a factor of about
2, but relative fluxes of strong lines
are estimated accurate to $\sim 10$ per cent.} 
\tablenotetext{c}{From Gaussian fits.}
\end{deluxetable}

\clearpage
\begin{deluxetable}{lrrrrc}
\footnotesize
\tablewidth{0pt}
\tablecaption{Fits to Radial Velocities}
\tablehead{
\colhead{Data set} & \colhead{$T_0$\tablenotemark{a}} & \colhead{$P$} &
\colhead{$K$} & \colhead{$\gamma$} & \colhead{$\sigma$}  \\
\colhead{} & \colhead{} &\colhead{(d)} & \colhead{(km s$^{-1}$)} &
\colhead{(km s$^{-1}$)} & \colhead{(km s$^{-1}$)}
}
\startdata
Absorption  & 2234.79927(8) & 0.044566(3) &  348(4) & 4(3) & 15 \\
H$\alpha$ emission & 2234.8239(9) & 0.04458(4) & 73(9)  
& $-78(6)$ & 38 \\
\enddata
\tablenotetext{a}{Blue-to-red crossing, HJD $- 2450000$.}
\end{deluxetable}

\clearpage


\begin{figure}
\plotone{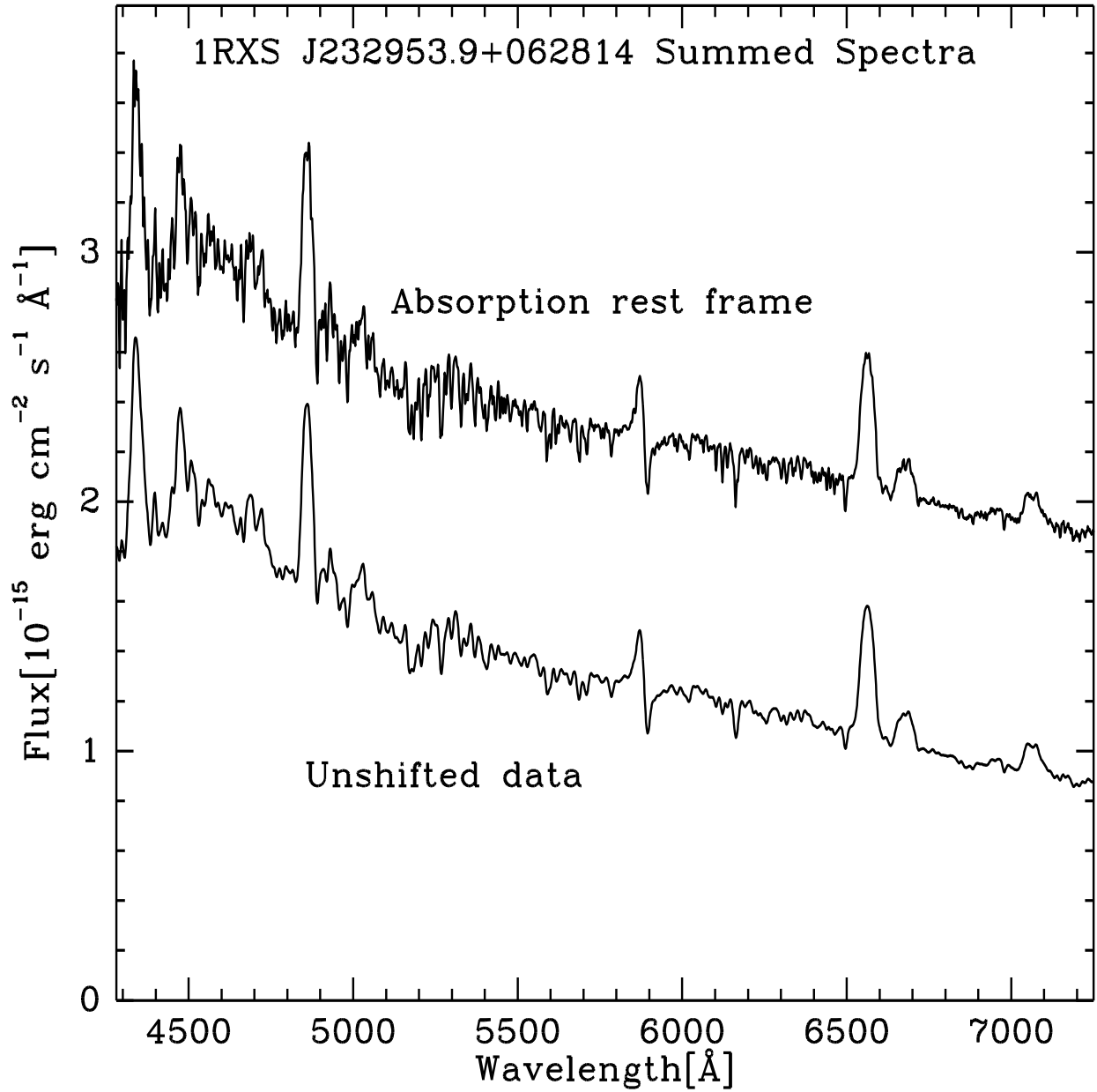}
\caption{Mean flux-calibrated spectrum.  In the upper trace,
the individual exposures have been shifted into the absorption
rest frame and the trace has been shifted upward.
}
\end{figure}

\clearpage

\begin{figure}
\plotone{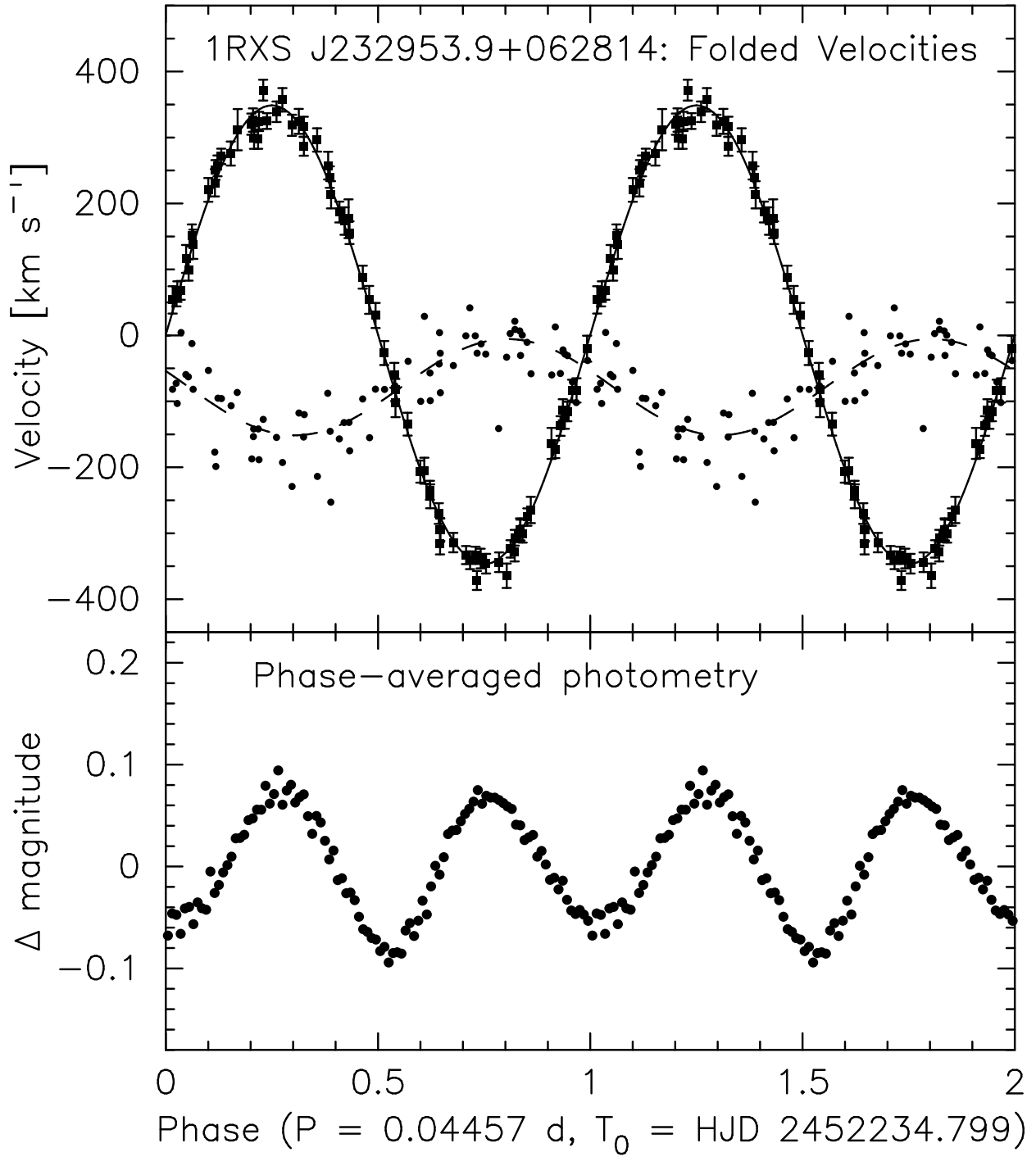}
\caption{{\it Upper panel:} Absorption (squares) and
emission (round dots) velocities folded on the binary period.
The best-fitting sinusoids are plotted.
{\it Upper panel:} Phase-averaged, unfiltered differential magnitudes.
All data are shown twice for continuity.
}
\end{figure}

\begin{figure}
\caption{Spectra rectified and shown as greyscale against phase.
In the lower panel the greyscale is chosen to emphasize the absorption
features, and in the upper it is chosen to show details of the
emission cores.  All data are shown twice for continuity.
}
\end{figure}

\begin{figure}
\plotone{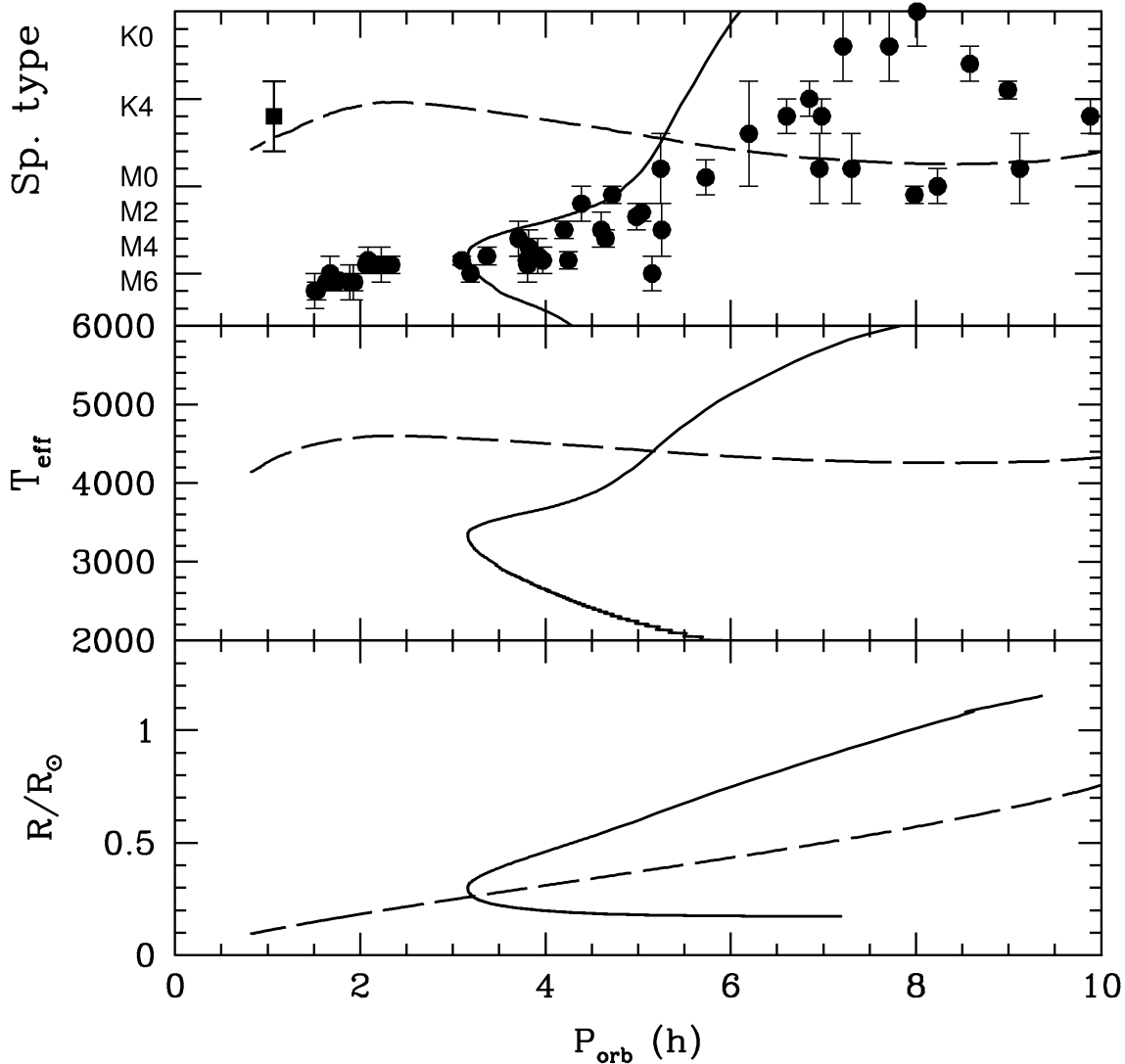}
\caption{Spectral type, effective temperature and radius versus orbital
period. Two evolutionary sequences with initial mass donor
$M_2$ = 1.2 $\msol$ and constant mass transfer 
$\dot M = 1.5 \times 10^{-9} \msolyr$ are displayed. The solid curve
corresponds to an initially unevolved donor and the dashed curve to
an initially nuclearly evolved donor starting mass transfer near the end
of central H burning phase. In the $P_{\rm orb}$ - spectral type diagram, 
the location of RX2329's secondary is indicated by a square. 
The other symbols (filled circles) are observations 
taken from \citet{beuermann98}.
}
\label{fig4}
\end{figure}

\end{document}